\documentclass[aps,preprintnumbers,nofootinbib,superscriptaddress,showpacs,letterpaper,twocolumn]{revtex4-1}
\pdfoutput=1
\usepackage{amsmath,amssymb,amsbsy}
\usepackage[utf8]{inputenc}
\usepackage{graphicx}
\usepackage[colorlinks=true, linkcolor=blue, filecolor=blue, urlcolor=blue, citecolor=blue, pdftex, plainpages=false]{hyperref}

\newcommand\buaff{Department of Physics and Center for Computational Science, Boston University, Boston, MA 02215, USA}
\newcommand\coaff{Department of Physics, University of Colorado, Boulder, CO 80309, USA}
\newcommand\Higgsaff{Higgs Centre for Theoretical Physics, School of Physics \& Astronomy, The University of Edinburgh, Edinburgh, EH9 3FD, UK}

\newcommand{\eq}[1]{Eq.~(\ref{#1})}
\newcommand{\fig}[1]{Fig.~\ref{#1}}
\newcommand{\Fig}[1]{Figure~\ref{#1}}

\newcommand{\refcite}[1]{Ref.~\cite{#1}}
\newcommand{\refcites}[1]{Refs.~\cite{#1}}

\begin{document}
\preprint{EDINBURGH 2015/31}

\title{Composite Higgs model at a conformal fixed point}
\author{R.C.~Brower}\affiliation\buaff
\author{A.~Hasenfratz}\affiliation\coaff
\author{C.~Rebbi}\affiliation\buaff
\author{E.~Weinberg}\affiliation\buaff
\author{O.~Witzel}\affiliation\Higgsaff
\date{\today}

\begin{abstract}
We propose to construct a chirally broken model based on the infrared  fixed point of a conformal system by raising the mass of some  flavors while keeping the others massless. In the infrared limit the massive fermions decouple and the massless fermions break chiral symmetry. The running coupling of this system ``walks"  and the energy range of walking can be tuned by the  mass of the heavy flavors. Renormalization group  considerations predict that the spectrum of such a system shows hyperscaling. 

We have studied a model with  four light and  eight heavy flavors coupled to SU(3) gauge fields and verified the above expectations. We determined the mass of  several hadronic states and  found that  some of them are in the 2-3~TeV range if the scale is set by the pseudoscalar decay constant $F_\pi \approx 250$~GeV.  The $0^{++}$ scalar state behaves very differently from the other hadronic states. In most of our simulations it is nearly degenerate with the pion and we estimate its mass  to be less than half of the vector resonance mass. 

\end{abstract}
\maketitle

\section*{Introduction}
Electroweak symmetry breaking (EWSB) and the nature of the Higgs boson  are  central questions of beyond the Standard Model (BSM) investigations. A gauge theory exhibiting spontaneous chiral symmetry breaking (S$\chi$SB) may describe EWSB when coupled to the Standard Model. In such a system three of the massless  Goldstone pions  become the longitudinal component of the $W^{\pm}$ and $Z$ bosons, while all other  hadronic states appear as  experimentally  observable excitations in the spectrum.  The physical energy scale is set by matching the decay constant of the pseudoscalar (pion) to the \emph{vacuum expectation value}\/ of the EWSB, i.e.~$F_\pi \approx 250$ GeV. BSM theories based on this construction are particularly interesting   as they predict several resonances around 2-3 TeV, an energy range  accessible at the LHC. The lightest vector meson state in our model is close to  2 TeV and could correspond to the recently reported    resonance \cite{Aad:2015owa,*CMS:2015gla}. 
These theories are based on similar concepts  originally introduced in the context of technicolor~\cite{Weinberg:1979bn,Susskind:1978ms,Eichten:1979ah,Appelquist:1991nm}.
Phenomenologically viable models must have properties quite different from QCD, suggesting they  are  likely  near the  conformal boundary.  Recent lattice simulations  with many fundamental flavors or  with fermions in higher representations have indeed revealed non-QCD-like properties~\cite{Bursa:2011ru,Aoki:2013zsa,Aoki:2014oha,Hietanen:2014xca,Fodor:2015vwa,Appelquist:2016viq}.

A composite BSM model with two massless fermions generates the required three Goldstone bosons. If the number of fermions  is larger  than two, as is the case in systems with fundamental flavors near the conformal boundary, the additional massless pseudoscalars have to acquire mass. While the precise mechanism of this could be complicated,  for an effective description one can simply add  a mass term to the  additional fermion flavors. In a model with $N_f$ fermions one would keep  $N_\ell = 2$ flavors massless and make $N_h = N_f - N_\ell$ fermions massive. That way,  the system will have only three massless Goldstone bosons in the infrared limit, yet the additional flavors  will have an influence on the spectrum.

When the total number of fermions increases above a critical value, the  system crosses the conformal boundary. The infrared properties are now characterized by a non-perturbative infrared fixed point (IRFP). 
Nevertheless the construction proposed above works just the same. Lifting the masses of all but $N_\ell=2$ flavors will lead to  S$\chi$SB  with three massless Goldstone bosons in the infrared limit. The presence of the conformal IRFP influences both the running of the gauge coupling and the spectrum. The idea to give mass to some of the flavors studied was previously discussed in Ref.~\cite{Dietrich:2006cm} and a similar construction, though with different phenomenology, has been proposed e.g.~in Refs. \cite{Vecchi:2015fma} and \cite{Luty:2004ye}.
 
In this paper we investigate the properties of such  a system, based on the   $N_f=12$ conformal model~\cite{Hasenfratz:2011xn,Cheng:2013xha,Cheng:2013eu,Itou:2014ota,Cheng:2014jba,Lombardo:2014pda}. We  lift the masses of $N_h=8$ fermions (heavy flavors)  and keep $N_\ell=4$ flavors light. This choice is  motivated by the lattice action we use in our simulations but  a chirally broken model with four light flavors  also has  phenomenological relevance.  An example is the  composite two Higgs-doublet model of \refcite{Ma:2015gra} that assumes four light flavors and  the Higgs bosons emerge as  pseudo-Goldstone states.  More commonly discussed models feature two massless fermions in the chiral limit and thus require simulations with $N_\ell=2$. Our choice is, however, sufficient to investigate general properties of mass split systems.
By changing $m_h$ from zero to $\infty$, our model interpolates between the conformal 12-flavor and the chirally broken 4-flavor systems.  If  $m_h > 0$,  chiral symmetry is spontaneously broken. In the next section, we  deduce that the hadron spectrum of the light  flavors shows hyperscaling in $m_h$  assuming $m_h$ is in the scaling regime of the IRFP and $m_\ell = 0$. Ratios of hadron masses  are independent of $m_h$, yet by tuning $m_h \to 0$ one can control the energy dependence  of  the gauge coupling, making it walking. Walking, in turn, can influence some of the infrared properties important to satisfy electroweak constraints.
 
 We study both the hadron spectrum and the running gauge coupling  in numerical simulations and  verify  theoretical  expectations. In particular  we show that ratios of hadron masses in the $m_\ell=0$ chiral limit are independent of $m_h$  in the scaling regime of the IRFP.  We compare the predicted spectrum with QCD and the conformal $N_f=12$ flavor  systems.  Despite  some similarities the spectrum we observe  is distinct from both QCD and the $N_f=12$ predictions. Setting the energy scale by $F_\pi \approx 250$~GeV we predict the lightest vector excitation  to be around 2 TeV.  The mass of the isomultiplet scalar is only slightly larger, while the nucleon and axial vector are both around 2.7 TeV.  At the fermion masses we can investigate, the mass of the isosinglet scalar is comparable to the pion. Thus predicting its chiral limit value is difficult and we only quote the bound $M_{0^{++}} /F_\pi \lesssim 4$. Simulations on larger volumes and smaller fermion masses are needed to make this prediction more precise. In addition, when coupled to the SM heavy quarks, radiative corrections will lower the mass of the scalar considerably~\cite{Foadi:2012bb}.  Preliminary results of both the spectrum and running coupling  were presented in \refcites{Brower:2014dfa,*Brower:2014ita,*Weinberg:2014ega,Hasenfratz:2015xca} and a publication with further details is in preparation \cite{Longpaper_inprep}.

\section*{Renormalization group structure and running coupling}
\begin{figure}[tb]
\centering
\hspace*{1.5mm}\includegraphics[width=0.915\columnwidth]{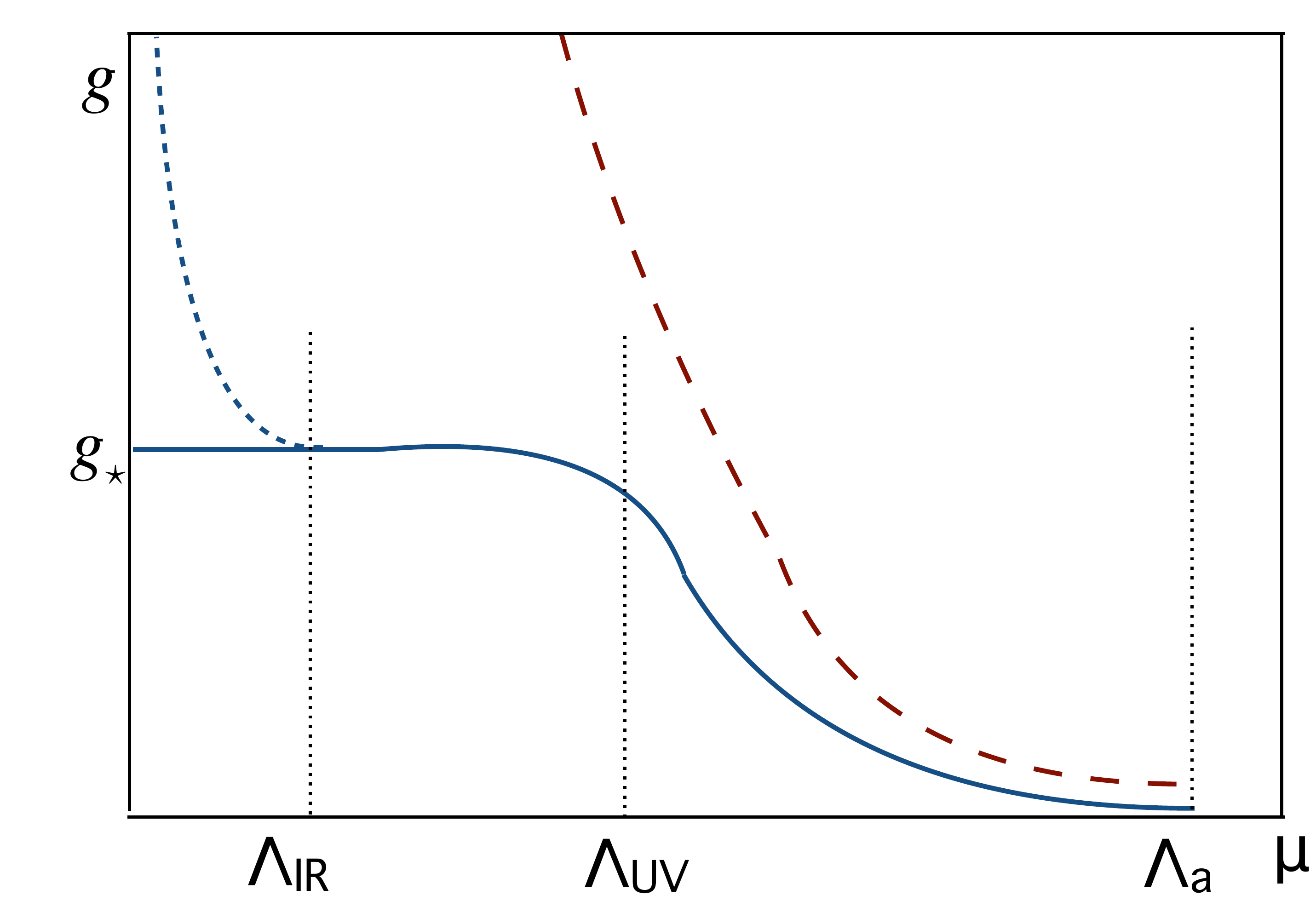}
\includegraphics[width=0.895\columnwidth]{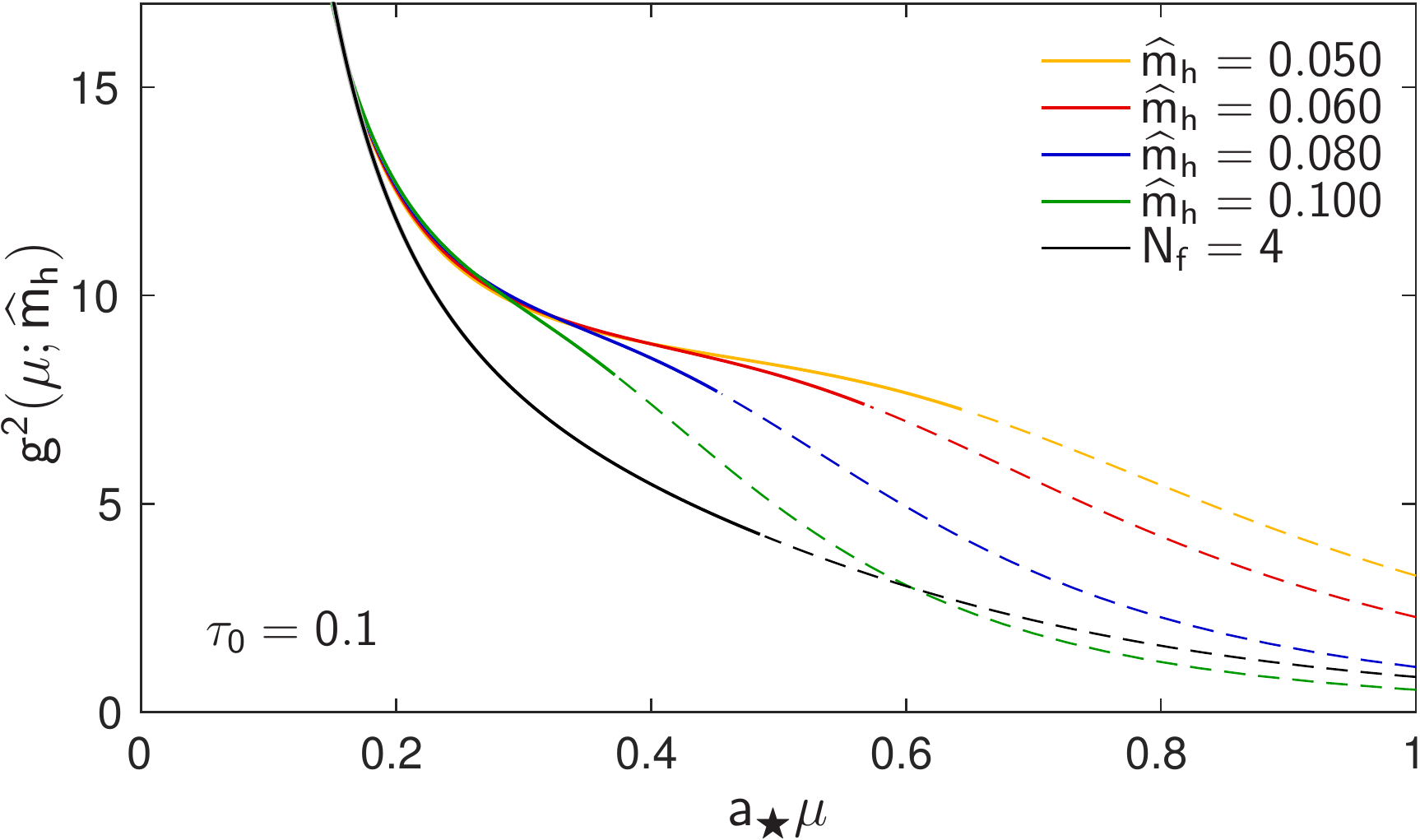}
\caption{Top: The expected running gauge coupling of conformal and mass-split systems. The solid blue curve sketches the evolution of the gauge coupling  in a conformal system. The dashed blue curve shows the change in a mass-split system while the red long-dashed curve describes  a situation where the fermions decouple before the gauge coupling could approach the conformal IRFP. Bottom:  Numerical results for the running coupling constant $g^2$  for different values of $\widehat m_h$ and in the $\widehat m_h=\infty$ (four-flavor limit) with $\widehat m_\ell$ extrapolated to the chiral limit. The emergence of the walking regime is evident as $\widehat m_h \to 0$. The dashed sections of the lines indicate where we  suspect cutoff effects  may be significant.   }
\label{fig:walking_coupling}
\end{figure}

Our 4+8 flavor model is built on the conformal IRFP of the SU(3) $N_f=12$ flavor system.  Near the IRFP, the only relevant parameter is the fermion mass. 
When all fermions have  a degenerate  mass  that  is  much smaller than the lattice cutoff $\Lambda_a \propto 1/a$, we expect hyperscaling with mass scaling dimension $y_m=1+\gamma_m \approx 1.25$~\cite{Hasenfratz:2011xn,Cheng:2013xha,Cheng:2014jba}.  The  Wilson  renormalization group (RG) equations predict that  a  change  in the energy scale $\mu \to \mu^\prime = \mu /b$ ($b>1$)  
 transforms the bare mass  $\widehat m = a m$ as $\widehat m \to \widehat m^\prime =  b^{y_{m}} \widehat m$, 
   while the bare gauge coupling
  $g \to g^\prime$ approaches its fixed point value $g_\star$. (For simplicity we consider only one gauge coupling, ignoring all other irrelevant couplings.)
It is straightforward to derive a scaling relation for any two-point correlation function  $C_H$~\cite{DeGrand:2009mt,DelDebbio:2010ze}
 \begin{align}
C_H(t; g, \widehat m_i,\mu) =  b^{-2y _H} C_H(t/b; g^\prime, \widehat m_i^\prime,\mu),
\label{eq:C_H2}
 \end{align} 
where  $y_H$ denotes the  scaling dimensions of operator $H$.  With repeated RG steps the irrelevant couplings approach the IRFP and the $b$ dependence of  \eq{eq:C_H2} can be written as $C_H \propto b^{-2y_H} F(t/b, b \widehat m^{1/y_m})$  where $F$ is some unknown function. Since correlation functions are expected to show an exponential decay $C_H(t) \propto exp(-M_H t)$ when the fermion mass is finite, \eq{eq:C_H2} implies the scaling relation
\begin{align}
a M_H \propto (\widehat m)^{1/y_m}.
\label{eq:M_scaling}
\end{align}
The renormalization group equation (\ref{eq:C_H2})  is valid even when some of the fermions are kept massless~\cite{Hasenfratz:2015xca}. The hyperscaling relation of \eq{eq:M_scaling}  remains unchanged with the  replacement $\widehat m \to \widehat m_h$,  and applies for both light and heavy flavored hadrons, as long as $\widehat m_\ell =0$ and all heavy flavors are degenerate.  Therefore dimensionless ratios, such as $M_H / F_\pi$ are independent of the heavy mass, as long as $\widehat m_h $ is in the scaling regime of the IRFP.  Since this system is chirally broken, even the  hadrons made up of light flavors are massive (except the Goldstone pions), and hyperscaling implies that their mass in units of $F_\pi$ is independent of the  heavy flavor mass.

\begin{figure*}[tb]
\centering
\parbox{0.085\textwidth}{\includegraphics[height=0.25\textheight]{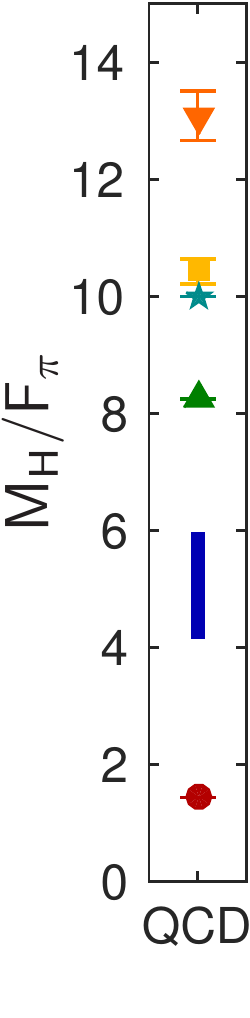}}
\parbox{0.20\textwidth}{\includegraphics[height=0.25\textheight]{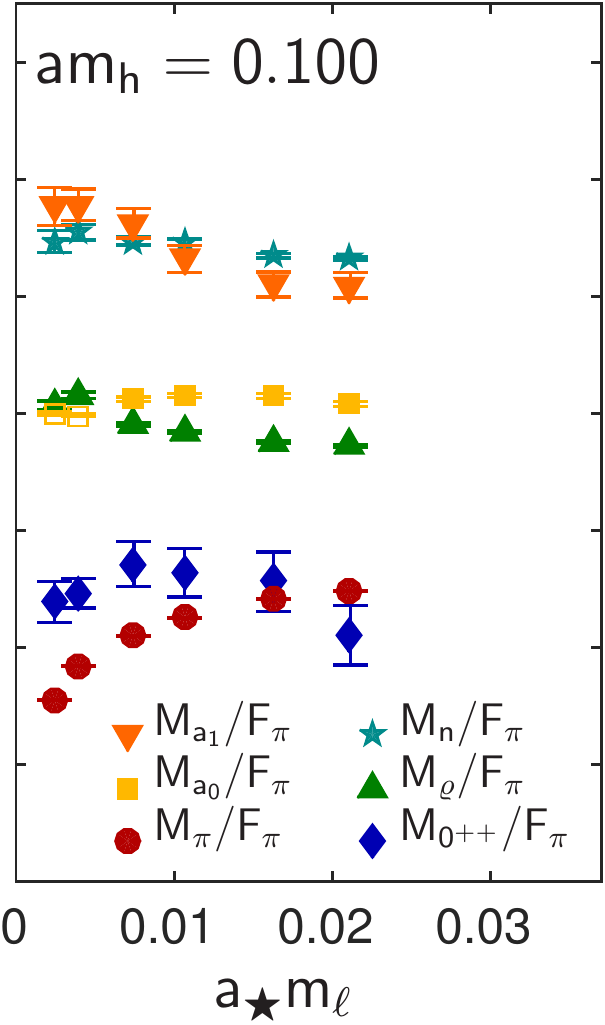}}
\parbox{0.20\textwidth}{\includegraphics[height=0.25\textheight]{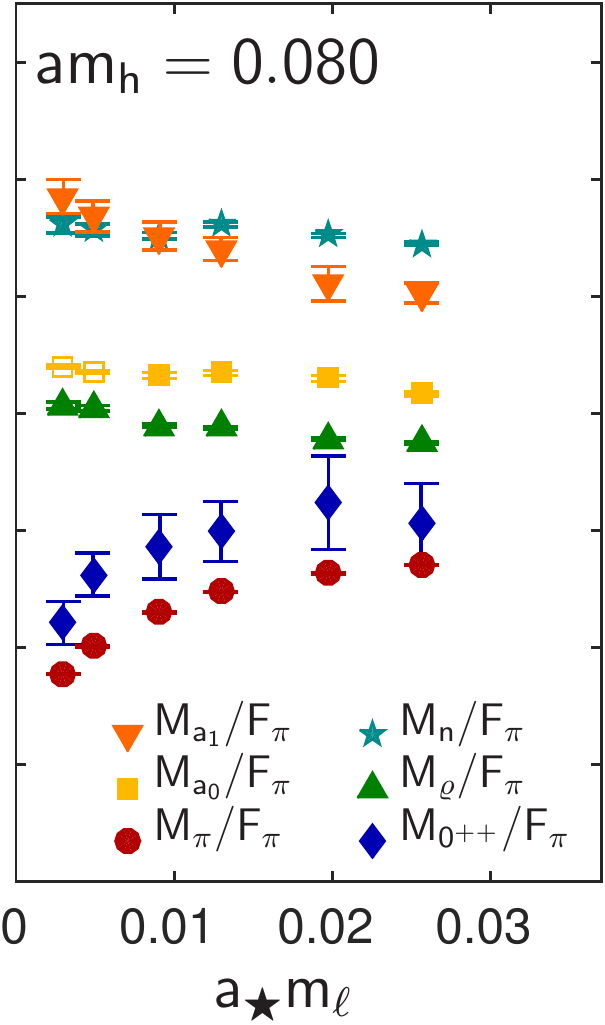}}
\parbox{0.20\textwidth}{\includegraphics[height=0.25\textheight]{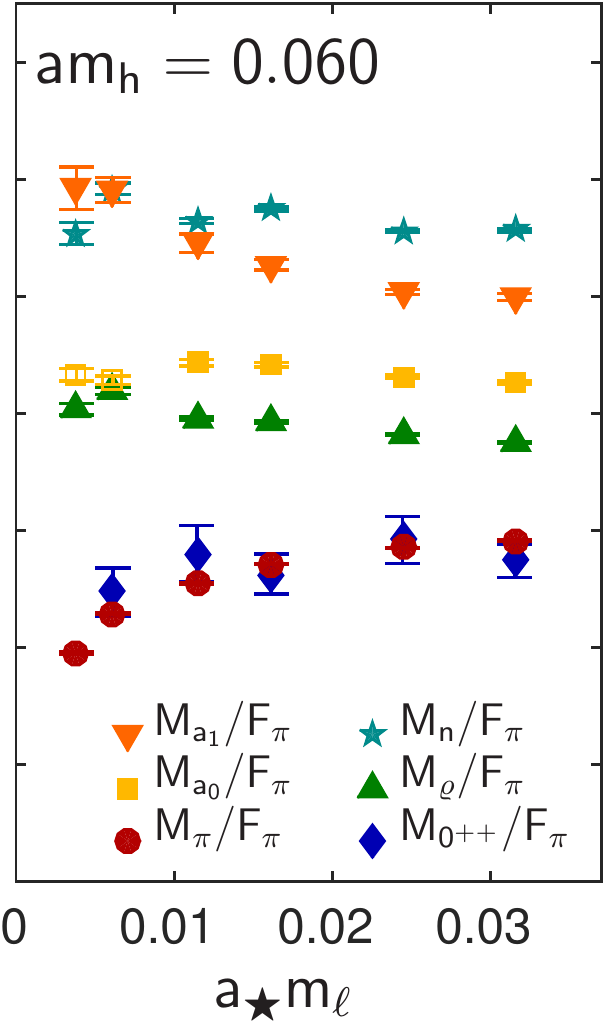}}
\parbox{0.20\textwidth}{\includegraphics[height=0.25\textheight]{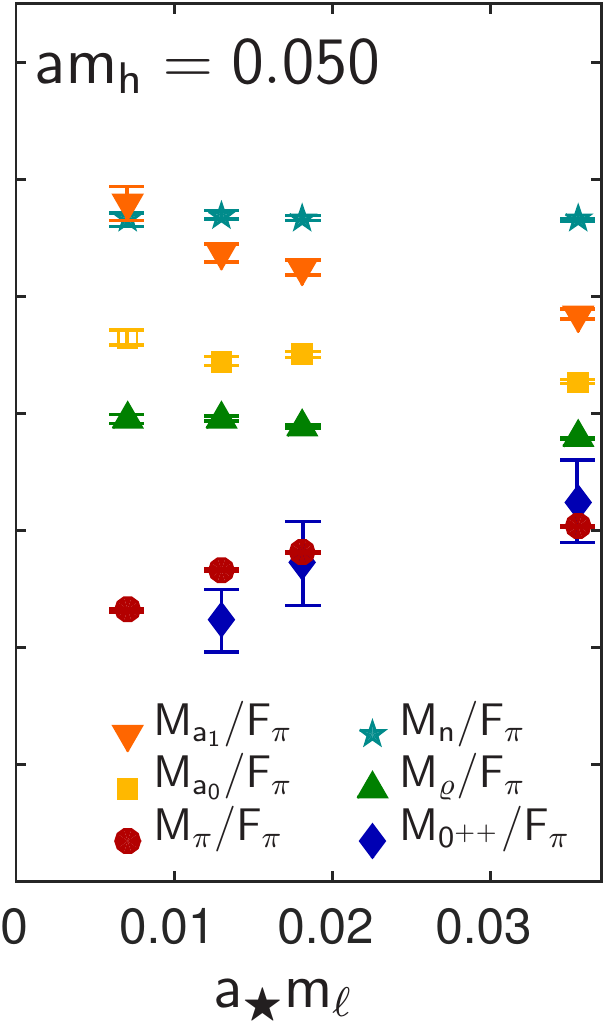}}
\parbox{0.051\textwidth}{\includegraphics[height=0.25\textheight]{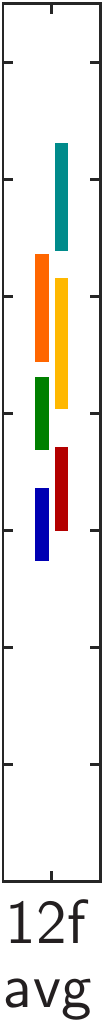}}
\caption{The pion, rho, isosinglet $0^{++}$ and isomultiplet  $a_0$ scalar, axial,  and nucleon mass of the light flavor spectrum in units of $F_\pi$. The first narrow panel shows the experimental values for QCD \cite{Agashe:2014kda} normalized by $F_\pi=94$ MeV, while the last one corresponds to average values obtained from $N_f=12$ flavor simulations~\cite{Aoki:2012eq,Fodor:2011tu,Cheng:2013xha,Aoki:2013zsa}. The four wider panels show the $N_f=4+8$ spectrum as the function of the light quark mass $a_\bigstar m_\ell$ for $am_h = 0.100$, $0.080$, $0.060$, and $0.050$.  If the chirally broken $N_f=4+8$  system   triggered EWSB,   $F_\pi \approx  250$ GeV would set the correct electroweak  scale. }
\label{fig:Spectrum}
\end{figure*}

Nevertheless  even for   $\widehat m_h \to 0$ the heavy flavors  will  influence other observables. Consider, e.g.,  the running gauge coupling as sketched in the top panel of \fig{fig:walking_coupling}. At high energies the coupling runs from the bare coupling toward the IRFP. At  the UV energy scale  denoted by $\Lambda_{UV}$   the gauge coupling reaches the vicinity of the  IRFP. Its value is close to $g_\star$ and  changes only slowly when further reducing the energy scale. In this regime the coupling ``walks.''  If all fermions were massless, $g(\mu \to 0) = g_\star$ as is indicated by the solid line in the figure. On the other hand if  some of the fermions are massive,  their mass becomes comparable to the cutoff at some energy scale, denoted by  $\Lambda_{IR}$,  and they decouple.  In this limit the system behaves  like a chirally broken  model with  $N_\ell$ massless fermions. The corresponding fast running coupling is denoted by the dashed blue lines in \fig{fig:walking_coupling}. The walking range between the scales $\Lambda_{UV}$ and $\Lambda_{IR}$ can be tuned by $\widehat m_h$, and a walking behavior in these systems is  guaranteed. 
The red  long-dashed curve in \fig{fig:walking_coupling} describes the case where the heavy  fermions decouple before the gauge coupling reaches the vicinity of the IRFP. This situation can be avoided by tuning $\widehat m_h \to 0$ and  is not considered here.

Our numerical simulations support the expectations outlined above. The bottom panel of \fig{fig:walking_coupling} shows the running coupling  calculated at five different values, $\widehat m_h= 0.050$, 0.060, 0.080, 0.100 and $\infty$ (i.e.~$N_f=4$). We define the energy dependent running coupling through the Wilson flow scheme and match the scales such that all five systems predict the same $g^2(\mu)$ in the infrared limit~\cite{Luscher:2010iy,Hasenfratz:2015xpa}. 
The $N_f=4$ system shows the expected  fast running, but a shoulder develops as $\widehat m_h$ is lowered. The dashed curves in the bottom panel of \fig{fig:walking_coupling} indicate regions where cutoff effects could be significant; however, theoretical considerations guarantee that the gauge coupling takes its IRFP value as $\widehat m_h \to 0$.  The similarity between the top and bottom panels of \fig{fig:walking_coupling} is striking and suggests that our simulations have entered the walking regime.  A walking gauge coupling leads to the enhancement of the fermion condensate and is necessary to satisfy electroweak  constraints.

\section*{Lattice simulations and  the hadron spectrum }
Wilson renormalization group considerations predict that the 4+8 flavor system shows   hyperscaling in the $a m_\ell=0$ chiral limit where dimensionless ratios of hadron masses are independent of the heavy mass $a m_h$. However, these ratios have to neither match the   $N_f=12$ nor the  $N_f=4$  flavor values.  In this section  we present numerical results for the hadron spectrum of the $N_f=4+8$ model at  four different $a m_h$ values. 

We use staggered fermions with nHYP smeared gauge links~\cite{Hasenfratz:2001hp,Hasenfratz:2007rf}  and a gauge action that is the combination of fundamental and adjoint plaquette terms. This  action has been used in  $N_f=12$ flavor simulations~\cite{Hasenfratz:2011xn,Cheng:2013xha,Cheng:2014jba} and we chose the parameters for this work  based on those results. We have carried out simulations at one gauge coupling, $\beta=4.0$, and four different values of the mass of the heavy flavors, $a m_h=0.050$, 0.060, 0.080 and 0.100. Based on the results of the finite size scaling study~\cite{Cheng:2014jba} we expect  that the three lightest values are within the scaling regime of the IRFP, while $m_h=0.100$ could be on the boundary.  We chose the  light fermion masses in the range $a m_\ell=0.003 - 0.035$ and the lattice volumes vary from $24^3\times48$ to $48^3\times96$. At many ($am_\ell,am_h$) mass values we consider two volumes to monitor finite volume effects.  We use the Wilson flow transformation to define the lattice scale~\cite{Luscher:2010iy}.  As $am_\ell \to 0 $ and $am_h\to 0$,  our simulations approach the $N_f=12$ conformal limit and consequently the lattice spacing decreases, requiring simulations on  increasingly larger volumes. Since we observe significant changes in the lattice spacing both when varying $am_h$ and $am_\ell$, we present our results in terms of a common reference scale $a_\bigstar$ that we define as the lattice scale on the ($36^3\times 64$, $am_h=0.080,am_\ell=0.003$) ensemble and convert results on other ensembles using ratios of the Wilson flow scale.

\Fig{fig:Spectrum} summarizes our results of the hadron spectrum. The  panels show  dimensionless ratios $M_H/F_\pi$ for   the pseudoscalar (pion), vector (rho), isomultiplet scalar ($a_0$), axial vector ($a_1$), nucleon ($N$) and isosinglet scalar ($0^{++}$) states. The first narrow panel shows the values  for QCD \cite{Agashe:2014kda} which is known to be similar to $N_f=4$, our $m_h \to \infty$ limit.  The last panel presents averages of numerical results for 12 degenerate flavors also accounting for the spread in the literature \cite{Aoki:2012eq,Fodor:2011tu,Cheng:2013xha,Aoki:2013zsa}. (In conformal systems the ratios are expected to be constant, up to corrections due to scaling violation.) The wider panels in the middle show the $N_f=4+8$ spectrum as  the   function of the light fermion mass measured in terms of the common lattice scale $a_\bigstar$  at our four $a m_h$ values.  The errors in \fig{fig:Spectrum} are statistical only. Based on the  comparison of different volumes, we  estimate that finite volume effects  are below the few percent level. We indicate data points by an open symbol where we suspect larger systematic effects.  Details will be discussed in \refcite{Longpaper_inprep}. 

Since the $N_f=4+8$ system is chirally broken,  we expect the pion mass to scale as $M_\pi \propto \sqrt{ m_\ell}$, while all other  hadrons should  acquire finite mass in the $am_\ell=0$ chiral limit. 
For large $a m_h$ in the limit  $a m_\ell \to 0$  we observe QCD-like behavior,  i.e. $M_\pi/F_\pi$ decreases, and $M_{a_1}/F_\pi$ increases toward the QCD value. On the other hand our system describes degenerate 12-flavors in the limit of $am_\ell=am_h \ll 1$.   At our largest $m_\ell$ values corresponding to  $am_\ell =0.035$ we find all six ratios to be in agreement with the 12-flavor averages.

  The ratios $M_H/F_\pi$ for the $\varrho$, $a_0$,  and nucleon states show a fairly linear dependence on $a_\bigstar m_\ell$, allowing for a simple   estimate  of their values  in to the $m_\ell=0$ chiral limit.  On the other hand extrapolating  the  isosinget $0^{++}$ state  to the chiral limit  is much more difficult. At $am_h=0.050$ and 0.060 the  $0^{++}$ is degenerate with the pion at all our $am_\ell$ values, indicating that our light fermions are not light enough to be in the chiral regime. For $am_h=0.080$ and 0.100 the $0^{++}$ slowly separates from the pion indicating a nonzero $M_{0^{++}}/F_\pi$ value as $m_\ell \to 0$. This  is the expected behavior because the $0^{++}$ is not a Goldstone boson.

\begin{figure}[tb]
\centering
\parbox{0.49\columnwidth}{\includegraphics[height=0.25\textheight]{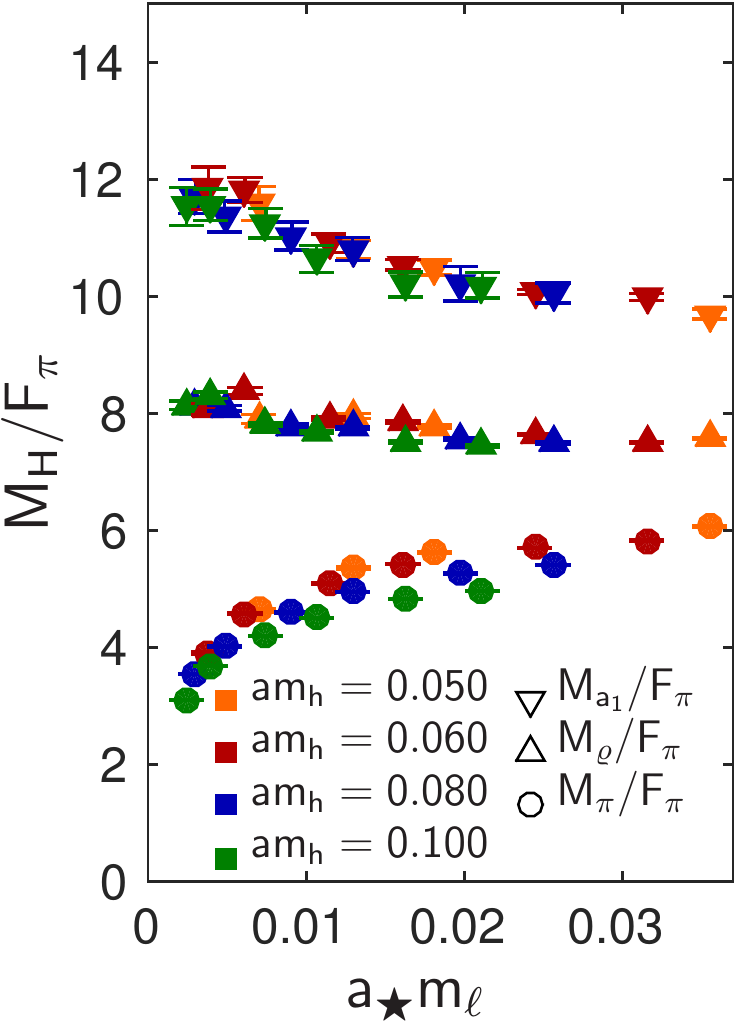}}
\parbox{0.40\columnwidth}{\includegraphics[height=0.25\textheight]{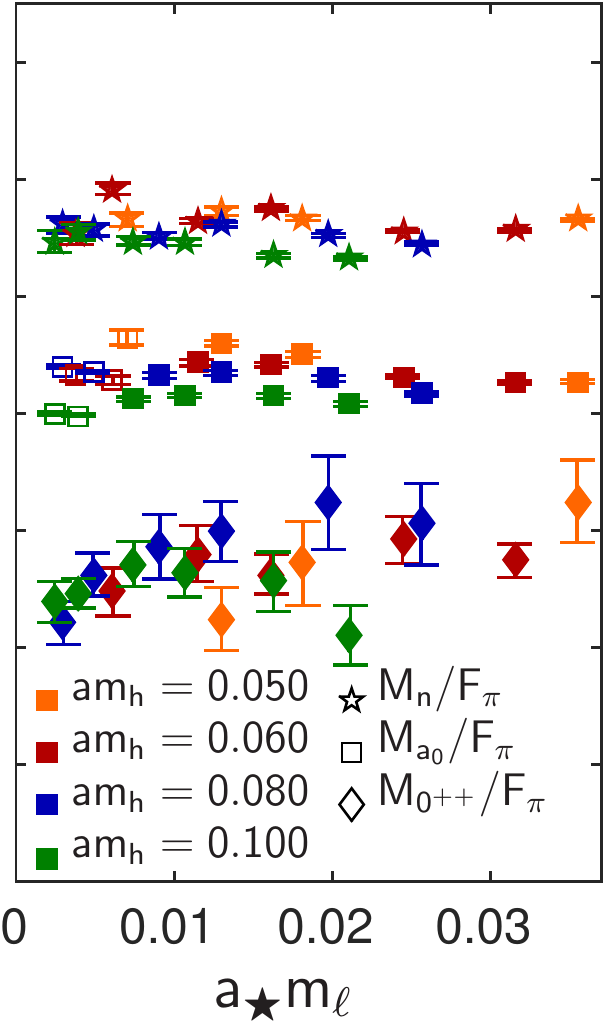}}
\caption{Combining all four $am_h$ values from \fig{fig:Spectrum}. The left panel shows the pion, rho, and  $a_1$, and the right panel shows the $0^{++}$, $a_0$, and nucleon states in units of $F_\pi$.}
\label{fig:ratios}
\end{figure}

Testing the hyperscaling hypothesis (valid only in the chiral limit), we compare our four different $am_h$ values in \fig{fig:ratios}. The left panel shows the ratios for the pion, rho, and $a_1$ exhibiting very little scatter in $am_h$. In particular $M_\varrho/F_\pi$ seems to be independent of the mass of the heavy flavors with a chiral limit of  just above 8.0. Thus the vector state in our model would be around 2 TeV if $F_\pi \approx 250$ GeV. 
Interestingly, many other near conformal models with an $SU(3)$ gauge group, like the $N_f = 8$ fundamental~\cite{Aoki:2013xza,Appelquist:2016viq}, the $N_f = 2$ sextet~\cite{Kuti2015}, and even the conformal $N_f = 12$ fundamental~\cite{Aoki:2012eq,Fodor:2011tu,Cheng:2013xha},  as well as  QCD predict almost the same $M_\varrho/F_\pi$ ratio.  Reference \cite{Appelquist:2016viq} argues that this could be a consequence of some remnant  vector meson dominance --- an idea worth exploring in the future. 

The right panel of \fig{fig:ratios} shows the ratios for the $0^{++}$, $a_0$, and nucleon. Here we observe   a larger spread for different $am_h$. At this point we cannot say whether this spread is due to systematical errors or a possible breakdown of hyperscaling, because corrections to hyperscaling may depend on the observable. In the case of the nucleon, the staggering of the different $am_h$ values might be interpreted   to approach the four-flavor  QCD-like limit. The $a_0$ state is more difficult to interpret, since systematic effects are increasing as $m_\ell \to 0$~\cite{Longpaper_inprep}. The $0^{++}$ receives contributions from disconnected diagrams, and thus the signal is noisier and the spread less significant. Overall, the $0^{++}$ shows a coherent trend: for larger $am_\ell$, it is degenerate with the pion --- a behavior previously observed in $N_f=8$ simulations~\cite{Aoki:2014oha,Rinaldi:2015axa,*Weinberg2015,Appelquist:2016viq} --- and approaches the value of the degenerate 12-flavor limit \cite{Aoki:2013zsa}. When $am_\ell$ decreases, the $0^{++}$ becomes lighter, and a linear extrapolation of the data for $a_\bigstar m_\ell \le 0.01$ would predict a chiral value $M_{0^{++}}/F_\pi \approx 4$. This value is half of the vector resonance and thus lower than the prediction in QCD. However it is possible that smaller $a m_\ell$ values, especially when $m_h$ is closer to the IRFP, could extrapolate to an even lower mass value. 
\section*{Outlook and Conclusion}
We study a model with spontaneously broken chiral symmetry built on the infrared fixed point of a conformal system by splitting the fermion masses. Specifically we study a system with $N_\ell=4$ light (massless) and $N_h=8$ heavy flavors. Renormalization group arguments imply this system exhibits a walking gauge coupling tunable with the heavy fermion mass and the spectrum shows hyperscaling.  Even though our model has four massless fermions in the infrared, its spectrum is not QCD-like. We predict  ratios $M_H/F_\pi$  for several hadronic states. Using the value $F_\pi \approx 250$~GeV,  the lightest vector excitation would  be around 2 TeV, the  mass of the isomultiplet scalar  only slightly larger, while the nucleon and axial vector are both around 2.7 TeV.  The  $0^{++}$ scalar state remains close to the pion in most of our simulations even when the rho is close to the two-pion threshold, a behavior not observed in QCD simulations. Thus, we expect the $0^{++}$ to be light, at most half of the vector state. However,  since our simulations imply  the light fermions are not yet in the chiral regime,  it is difficult to predict the mass of the scalar state.  In addition the scalar mass will decrease further due to top-loop corrections when coupled to the SM. 

In summary our model exhibits a light $0^{++}$ state (Higgs candidate) and predicts additional resonances in the 2-3 TeV range. Most remarkably the latter predictions seem to be rather universal for BSM models based on the SU(3) gauge theory and are most likely within the reach LHC run II. If the confidence on the 2 TeV resonance \cite{Aad:2015owa,*CMS:2015gla} increases, further studies  are warranted.  We are considering simulations of models with only two light flavors that are closer to the conformal window.


\begin{acknowledgments}
The authors thank their colleagues in the LSD Collaboration for fruitful and inspiring discussions. 
Computations for this work were carried out in part on facilities of the USQCD Collaboration, which are funded by the Office of Science of the U.S.~Department of Energy, on computers at the MGHPCC, in part funded by the National Science Foundation, and on computers allocated under the NSF Xsede program to the project TG-PHY120002. 
We thank Boston University, Fermilab, the NSF and the U.S.~DOE for providing the facilities essential for the completion of this work. R.C.B., C.R.~and E.W.~were supported by DOE grant DE-SC0010025 and in addition acknowledge the support of NSF grant OCI-0749300. A.H.~acknowledges support by the DOE grant DE-SC0010005. O.W.~is supported by STFC, grant ST/L000458/1. This project has received funding from the European Union's Horizon 2020 research and innovation programme under the Marie Sk{\l}odowska-Curie grant agreement No 659322. R.C.B., A.H. and C.R.~thank the Aspen Center for Physics, which is supported by National Science Foundation grant PHY-1066293. R.C.B., A.H., and~O.W.~thank the KITP, Santa Barbara, supported in part by the National Science Foundation under Grant No.~NSF PHY11-25915. A.H.~and~O.W.~thank the  CERN Theory group for their hospitality during the completion of this manuscript.
\end{acknowledgments}

\bibliography{../General/BSM}
\bibliographystyle{apsrev4-1} 
\end{document}